# Temperature-activated layer-breathing vibrations in few-layer graphene

Chun Hung Lui,[*,†,⊥] Zhipeng Ye,[‡,⊥] Courtney Keiser,[‡] Xun Xiao,[‡,∥] and Rui He[*,‡]

[†] Department of Physics, Massachusetts Institute of Technology, Cambridge, Massachusetts 02139, USA
[‡] Department of Physics, University of Northern Iowa, Cedar Falls, Iowa 50614, USA
[∥] College of Physical Science and Technology, Yangtze University, Jingzhou, Hubei 434023, China

[⊥] These authors contributed equally to this work.
*Corresponding authors (email: lui@mit.edu; rui.he@uni.edu)

**ABSTRACT:** We investigated the low-frequency Raman spectra of freestanding few-layer graphene (FLG) at varying temperatures (400 – 900 K) controlled by laser heating. At high temperature, we observed the fundamental Raman mode for the lowest-frequency branch of rigid-plane layer-breathing mode (LBM) vibration. The mode frequency redshifts dramatically from 81 cm$^{-1}$ for bilayer to 23 cm$^{-1}$ for 8-layer. The thickness dependence is well described by a simple model of coupled oscillators. Notably, the LBM Raman response is unobservable at room temperature, and it is turned on at higher temperature (>600 K) with a steep increase of Raman intensity. The observation suggests that the LBM vibration is strongly suppressed by molecules adsorbed on the graphene surface, but is activated as desorption occurs at high temperature.

**KEYWORDS:** Layer breathing mode, shear mode, few layer, graphene, Raman, adsorption

Two-dimensional (2D) graphene materials have attracted much attention due to their diverse physical properties. In particular, their atomic thickness gives rise to high surface-to-volume ratio and hence great sensitivity to environmental influences. For instance, the surface adsorbates are known to induce significant charge doping in graphene, modulating the properties of the electrons [1-2] and also, indirectly, the properties of some phonons through their coupling to the electronic system [3]. These adsorbates are also expected to mechanically perturb the lattice vibrations. This latter effect has, however, not been revealed in the widely-studied intralayer phonons as well as the interlayer shear (C) mode phonons [4-9], presumably due to the weak coupling between the physisorbed molecules and the lateral atomic motions of these vibrations [5]. Few-layer graphene (FLG), however, possesses a special type of vibrations: the layer breathing modes (LBMs, also called ZO' modes) with perpendicular displacement of the graphene layers (Figure 1). These out-of-plane interlayer modes are expected to be particularly sensitive to external perturbations. Investigation of the LBMs and their environmental sensitivity is therefore a subject of great scientific and practical significance.

There have been extensive theoretical [10-15] and experimental studies on the interlayer vibrations of FLG [4-9, 16-18] and other 2D layered crystals, e.g. MoS$_2$ and WSe$_2$ [19-23]. However, direct observation of the LBMs in FLG is challenging because they couple weakly to light and are probably susceptible to the external influence. Indeed, the LBM is optically silent in graphite [7, 10]. Although the LBMs may become Raman active in FLG, they have thus far only been observed indirectly through the doubly resonant combination and overtone Raman modes (LOZO' and 2ZO' modes) [16-18]. While these higher-order Raman processes effectively up-shift the mode frequency and enhance the Raman response through the electronic resonance, they obscure the intrinsic properties of the LBMs. First, the involvement of other phonons, such as the LO phonons in the LOZO' combination mode, prevent us from accurately extracting the LBM frequency. Second, the double-resonance Raman process only provides access to phonons with finite momentum, but not the fundamental rigid-plane LBMs at the Brillouin zone center. Third, among the numerous LBM branches in FLG, the previous studies could only reveal the branches with relatively high frequency (> 50 cm$^{-1}$). Therefore, it is desirable to have direct access to the lowest-frequency rigid-plane LBM, which has the simplest vibration pattern, highest symmetry, and supposedly the strongest environmental sensitivity.

In this Letter, we report the direct observation of the lowest-frequency rigid-plane LBM optical phonons in FLG with layer number $N$ from 2 to 8. By using sensitive Raman probe with frequency range accessible down to 10 cm$^{-1}$ on freestanding FLG samples, we discovered a new sharp Raman line in FLG. This Raman feature redshifts dramatically from 81 to 23 cm$^{-1}$ as the number of layers increases from 2 to 8, in sharp contrast with the Raman shear (C) mode that blueshifts with increasing layer number. The measured dependence of the mode frequency on the layer number matches the prediction of the lowest LBM branch based on a linear-chain model. Remarkably, the observed LBM exhibits strong variation with the graphene temperature (400 – 900 K) controlled by laser heating. Its Raman response vanishes at room temperature but is turned on at elevated temperature (>600 K). The observation suggests that adsorbates on the graphene surface suppress the LBM vibration at room temperature, and desorption of these surface molecules activates the LBM at high temperature. Our findings provide important evidences of the direct influence of the surface adsorbates on the lattice vibrations of graphene.

We investigated freestanding Bernal-stacked graphene samples with layer number $N$ = 1 to 8 layers. These samples were prepared by micro-mechanical cleavage of kish graphite on quartz substrates with pre-patterned trenches (4-μm-wide, 1-μm-deep). The graphene areas suspended over the trenches are isolated from any disturbance of the substrate. The number of layers and stacking order of the FLG samples were determined by their optical contrast, infrared [24-27] and Raman spectra [16-18, 28]. In particular, by using the Raman features of the LOZO' combination mode, 2ZO' overtone mode and 2D mode, we can determine the number of layers with monolayer accuracy and the stacking order (AB or ABC stacking) of our samples [16-18, 28]. We performed Raman measurements at room temperature, using a commercial Horiba Labram HR micro-Raman system equipped with a 532-nm excitation laser, a 100 × objective lens, a 1800-groove/mm grating, and a thermoelectric-cooled charge-coupled device (CCD). The Raman setup offers access to frequencies down to 10 cm$^{-1}$ and spectral resolution of 0.5 cm$^{-1}$. We conducted the measurements with laser power ($P$) from 1 to 10 mW in an argon-purged environment to reduce the background signal from the air.

We first present the Raman spectra under high excitation laser power ($P \sim 9$ mW). The laser heating was found to cause a ~3 cm$^{-1}$ redshift of frequency of the interlayer modes. No degradation of the FLG samples was found as revealed from the negligible D Raman band (see Supporting Information). As we will discuss later, the laser heating is crucial to activate the layer-breathing vibrations. Figure 1 displays the Raman spectra of monolayer graphene (1LG) and bilayer graphene (2LG) in the frequency range of 15 – 225 cm$^{-1}$. While the 1LG spectrum is featureless, the 2LG spectrum exhibits a few Raman peaks at 30, 81, 113, 174 and 198 cm$^{-1}$.

The absence of these features in 1LG indicates that they all originate from the interlayer vibrations in 2LG. Previous studies have identified the prominent peak at 30 cm$^{-1}$ (marked by a blue dot in Figure 1) as the C mode arising from the in-plane shear-mode vibration [4-9], and the double peaks at 170 – 200 cm$^{-1}$ (cyan dots) as the 2ZO' overtone mode arising from the emission of two LBM phonons with opposite finite momenta through intra-valley double-resonance Raman processes [18].

The peak at 81 cm$^{-1}$ (red dot) is observed for the first time. There are evidences that it is the fundamental Raman mode for the rigid-plane LBM in 2LG. First, the relatively narrow line width (4.5 cm$^{-1}$) of this peak implies that it does not arise from any double-resonance process, which typically leads to broad line width (>10 cm$^{-1}$) due to the emission of phonons with different momenta. The new peak at 81 cm$^{-1}$ should therefore be associated with phonons at the Brillouin zone center. Second, 2LG has relatively simple structure. The zone-center interlayer vibrations comprise only one doubly-degenerate shear mode and one LBM. Since the shear mode has been identified, the new Raman feature must come from the LBM. Third, the frequency of this new peak matches the ZO' mode frequency predicted by theory [10, 14] and determined experimentally from the doubly resonant 2ZO' overtone mode [18] (cyan dots in Figure 1). Our prior study of the 2ZO' modes [18], which was also conducted with high excitation laser power (a few mW) on suspended FLG samples, has shown that the 2ZO' mode blueshifts with increasing excitation photon energy and concomitant increase of phonon momentum. By extrapolating the 2ZO' mode frequency to zero phonon momentum, the ZO' phonon frequency was estimated to be 80 ± 2 cm$^{-1}$ at the zone center. This value agrees well with the frequency of the new peak (81 cm$^{-1}$). Therefore, we conclude that the new Raman mode at 81 cm$^{-1}$ arises from the zone-center ZO' phonon.

The Raman peak at 113 cm$^{-1}$ (green dot in Figure 1) is weaker than the C mode but stronger than the ZO' mode. Its narrow line width (~3 cm$^{-1}$) implies that it does not arise from any double-resonance Raman process. As the frequency of this peak is very close to the sum of the C-mode and ZO'-mode frequencies, we here assign it as the C+ZO' combination mode [8].

As the LBM phonons are highly sensitive to interlayer interactions, it is interesting to explore their behavior in FLG with greater layer thickness. An $N$-layer graphene system possesses totally $N$-1 ZO' phonon branches, denoted here as ZO'$_N^{(n)}$, where the superscript $n = 1, 2, … N$-1 denotes branches from high to low frequency. These numerous ZO' branches are expected to give rise to complex Raman spectra in thicker FLG samples, as shown by the previous studies of LOZO' and 2ZO' modes [17-18]. We have performed Raman measurements on suspended FLG samples with layer number $N$ from 2 to 8 and bulk graphite. Figure 2 displays their spectra in the frequency range of 18 – 95 cm$^{-1}$. To our surprise, we observe rather simple spectra for all the FLG samples: there are only two noticeable peaks for each spectrum. One of them is the C mode, which blueshifts from 30 to 40 cm$^{-1}$ as the layer number increases (blue symbols in Figures 2 and 3). According to prior studies[4-9], this C Raman mode corresponds to the highest branch of the shear-mode phonon, whose frequency increases asymptotically towards the bulk value as $N$ increases. Other lower shear branches have much weaker Raman response. In contrast, the other group of peaks (red dots in Figures 2 and 3) exhibits a completely different behavior from the C mode. It redshifts dramatically from 81 cm$^{-1}$ in 2LG to 23 cm$^{-1}$ in 8LG and disappears in graphite. The observation suggests that this second group of peaks does not arise from the highest LBM branch, as one might expect.

To identify the origin of the new set of Raman peaks, we have calculated the frequencies of all ZO'$_N^{(n)}$ modes in FLG based on a coupled-oscillator model with only nearest-layer coupling. In this simple model, the layers are treated as a linear chain of $N$ masses connected by constant springs. The predicted frequency of the ZO'$_N^{(n)}$ normal modes can be expressed in a simple analytical form[29]:

$$\omega_N^{(n)} = \omega_o \, cos \, (n\pi/2N). \qquad (1)$$

Here $\omega_o$ is the LBM frequency of bulk graphite at the zone center. From our measured ZO'$_2^{(1)}$ frequency in 2LG, $\omega_2^{(1)} = \omega_o/\sqrt{2} = 81$ cm$^{-1}$, we obtain $\omega_o = 114.6$ cm$^{-1}$. By comparing our experiment with the theory, we find that the calculated frequency of the lowest branch (the ZO'$_N^{(n)}$ modes with $n = N$-1) matches exactly the frequencies of the new group of peaks (Figure 2 and 3). We therefore assign them as the ZO'$_N^{(N-1)}$ modes.

Our observation indicates that the lowest LBM branch dominates the Raman response, with no observable Raman lines arising from other higher LBM branches. We can understand this Raman behavior from the atomic displacement of these normal modes. The lowest branch ZO'$_N^{(N-1)}$ mode has a remarkably simple vibration pattern, in which the layers contract and expand with respect to the central plane of the structure (Figure 3a). In this vibration, the similar motion of adjacent layers results in low restoring force and thus the lowest fundamental frequency. Additionally, the vibration induces the largest compression on FLG, giving rise to the highest variation of polarizability and hence the strongest Raman response[30]. In contrast, the higher-frequency LBM branches have layer displacements that compensate one another, reducing partially or fully the change of polarizability and hence the Raman response.

We can further understand the Raman behavior by symmetry analysis based on group theory [10-11]. Bernal-stacked FLG with even (odd) number of layers has inversion (mirror) crystal symmetry. The ZO'$_N^{(n)}$ modes have even parity under inversion (mirror reflection) with $A_{1g}$ ($A_1'$) representation when $N - n$ is an odd mode number. They are Raman active. However, the ZO'$_N^{(n)}$ modes have odd parity and are Raman forbidden when $N - n$ is an even number. Therefore, the lowest ZO' branch is Raman active and the second lowest ZO' branch is Raman forbidden for all FLGs, while the higher branches might be Raman active or inactive. Similar LBM symmetry properties and overall decrease of Raman activity in higher LBM branches have also been observed in FLG with rhombohedral (ABC) stacking order [31] and in layered dichalcogenide crystals [20-23]. It is worthwhile to note that the Raman mechanism and selection rules for these fundamental modes are different from those for the overtone/combination modes that are produced by the double-resonance Raman processes. Prior studies showed that LBMs of all branches are observable in the 2ZO' overtone modes [18] and the LOZO' combination modes [17]. However, here only the lowest ZO' branch is observed in the first-order fundamental Raman mode.

The fundamental LBM behavior is very different from that of the shear (C) modes. Due to the AB stacking in our FLG samples, the highest shear branch, with oppositely directed lateral motion of all pairs of adjacent layers, produces the largest change of polarizability and hence the strongest Raman response among all the shear branches. In the group-theory analysis [10-11], the highest shear branch has the $E_g$ symmetry and is Raman active for all FLGs, whereas the other lower branches might be Raman active or inactive. (Detailed comparison of the shear modes and LBMs can be found in Table S1 and S2 of the Supporting Information.)

In addition to the shift of mode frequency, the lowest-frequency ZO' mode also exhibits interesting thickness dependence in their Raman line width. The full width at half maximum (FWHM) of the ZO'$_N^{(N-1)}$ mode decreases from 4.5 to 0.8 cm$^{-1}$ as $N$ increases from 2 to 8 (Figure 3b). In FLG, the Raman line width of phonons mainly arises from their decay into other lower-energy phonons or electron-hole pairs. The decay rates of both processes are expected to increase with the phonon energy. The observed decrease of ZO'$_N^{(N-1)}$ mode line width therefore reflects the corresponding decrease of ZO'$_N^{(N-1)}$ phonon energy with increasing $N$ (Figure 3b). This interpretation is also consistent with the opposite dependence of the C-mode line width on $N$ since the C-mode phonon energy increases with increasing $N$ (Figure 3b).

The out-of-plane layer vibration of LBM in atomically thin graphene is highly sensitive to external perturbations, such as the presence of a substrate or adsorbates. Indeed, the fundamental layer-breathing Raman mode was unobservable in FLG samples deposited on substrates [4-9]. To address the role of environmental effect, we have measured the suspended FLG samples with varying lattice temperature controlled by the excitation laser power. We found that the LBM only appeared in suspended FLG samples under high excitation power. Figure 4a displays the Raman spectra of a suspended 8LG sample with incident laser power from 1 to 9.5 mW. The spectra clearly show the onset of LBM at 5 mW. We have also measured the Stokes and anti-Stokes G Raman modes and estimated the graphene temperature (with an uncertainty of a few tens of degrees) from their intensity ratio (Figures 4b and c). Due to the low rate of heat dissipation in suspended FLG, the local lattice temperature could reach ~900 K at high excitation power.

Figure 4d displays the integrated Raman intensities (normalized by the excitation power) of the ZO', C and G modes as a function of the laser power (bottom axis) and temperature (top axis). Both the C and G modes exhibit rather constant Raman response at all laser powers, indicating that their Raman activity is not affected by the temperature. In contrast, the LBM Raman response shows strong variation with the temperature. The LBM signal vanishes at low excitation power ($P < 5$ mW). It is turned on with a steep increase of Raman intensity at $P \sim 5$ mW ($T \sim 620$ K), and the intensity continues to grow with the increase of laser power (temperature). The observed change of LBM signal was reversible as the same spectra were seen when we decreased the laser power from high to low values. We note that the ZO' phonon population only varies moderately in the measured temperature range (dashed line in Figure 4d) and thus cannot account for the steep change of its Raman response.

Here we propose an explanation for the strong dependence of the ZO' mode intensity on the laser power and lattice temperature. In our experiment, there is a non-negligible fraction of air molecules, such as $N_2$, $O_2$ and $H_2O$, even in the argon-purged environment. These residue air molecules can easily adsorb on the graphene surface. In particular, recent theoretical calculations [32-36] predicted that $H_2O$ has adsorption energy of 40 – 150 meV on graphene, well above the room-temperature thermal energy but comparable to that of the onset temperatures determined in our experiment (>50 meV). These adsorbates can randomly modify the LBM frequency, distort and frustrate the out-of-plane layer vibrations, and thereby suppress the ZO' mode Raman response. As the graphene temperature increases, however, these molecules start desorbing from the graphene surface, and the ZO' mode vibrations are activated. With further increase of the temperature, more and more adsorbates and those with higher binding energies continue to desorb from the graphene surface, further recovering the ZO' vibrations. A reversible process with no hysteresis happens when the laser power decreases from high to low values because the adsorption and desorption events are expected to occur in a time scale much shorter than that needed to carry out the Raman measurement (~20 minutes per spectrum).

As FLGs with lower number of layers are more susceptible to surface adsorbates, we expect higher activation temperatures in the thinner samples. We have examined suspended 6LG and 4LG samples and found that the LBM onset temperatures were ~670 K and ~710 K, respectively (see Figures S2 and S3 in the Supporting Information), both higher than that of 8LG (~620 K). The LBM of 2LG remained relatively weak up to ~900 K (Figure 2), the highest achievable temperature in our experiment. This implies a highest activation temperature (>900 K) for 2LG among all FLGs. The observed general increase of activation temperature in thinner graphene layers further supports the adsorption mechanism presented above. In addition, we also observed the thermal activation of the doubly resonant overtone 2ZO' modes[18], but with somewhat lower onset temperature (see, for example, the result of 4LG in Figure S4 of the Supporting Information). The finding suggests that the influence of the adsorbates can be reduced to some extent by electronically resonant Raman process. The thermal activation effect was found to be much weaker in the LOZO' combination mode[17], presumably due to the involvement of LO phonons, which are highly Raman active and insensitive to surface adsorbates. Further investigation is needed to better understand the influence of these additional factors on the thermal activation of LBMs.

The C and G modes vary only slightly with temperature (Figure 4d). This indicates that these vibrations, with tangential atomic displacements, are much less sensitive to the adsorbates. This finding is consistent with a previous study[5] that showed small variation of the shear mode for suspended and supported FLG samples and for samples covered with a dielectric top layer. We also remark that such strong frustration of LBM by substrates or adsorbates was not seen in atomically thin layered transition metal dichalcogenides[19-23]. This could be due to weaker absorption processes, larger mass density or complex monolayer structure in these materials. For example, the unit cell of $MoS_2$ and $WSe_2$ is, respectively, ~7 and ~14 times heavier than that of graphene. The motion of these heavier atoms is expected to be less susceptible to the presence of adsorbates. In addition, each monolayer of $MoS_2$ ($WSe_2$) has the S-Mo-S (Se-W-Se) internal stacking structure, with the center of mass relatively far from the adsorbates (or substrates), thus reducing the influence of external perturbations. These differences between graphene and transition metal dichalcogenides suggest that the unique environmental sensitivity of LBMs in FLG originates from the atomically-thin thickness and lightweight characteristics of the carbon layers.

In conclusion, we discovered the fundamental Raman mode for the lowest-frequency layer-breathing mode (LBM) vibration in freestanding few-layer graphene (FLG). We determined the LBM frequency and linewidth, and explained their evolution with the number of layers. These results are useful to verify and refine the theories on the lattice dynamics of graphene-based materials. In addition, we found that the LBM is strongly suppressed at room temperature, but shows an onset of Raman response at elevated temperature. We attribute the suppression of LBM to mechanical damping due to surface adsorbates at room temperature, and the activation of LBM to desorption of surface molecules at high temperature. Our research can be further confirmed and extended by investigating the LBM in vacuum conditions and with controlled amount of selected gases. Such studies will help us understand the adsorption processes of different gases on graphene surface as well as develop LBM-based mass detectors that may complement the existing gas sensors using electrical detection methods[1-2].

We thank D. Boschetto, E. B. Barros and M. S. Dresselhaus for helpful discussion. Acknowledgment is made to the Donors of the American Chemical Society Petroleum Research Fund (Grant 53401-UNI10) and the National Science Foundation MRI Grant (No. DMR-1337207) for support of this research. R. H. acknowledges support from UNI Faculty Summer Fellowship. X. X. acknowledges support from The Basic Sciences Research and Development Fund of Yangtze University under grant No. 2013cjp15.

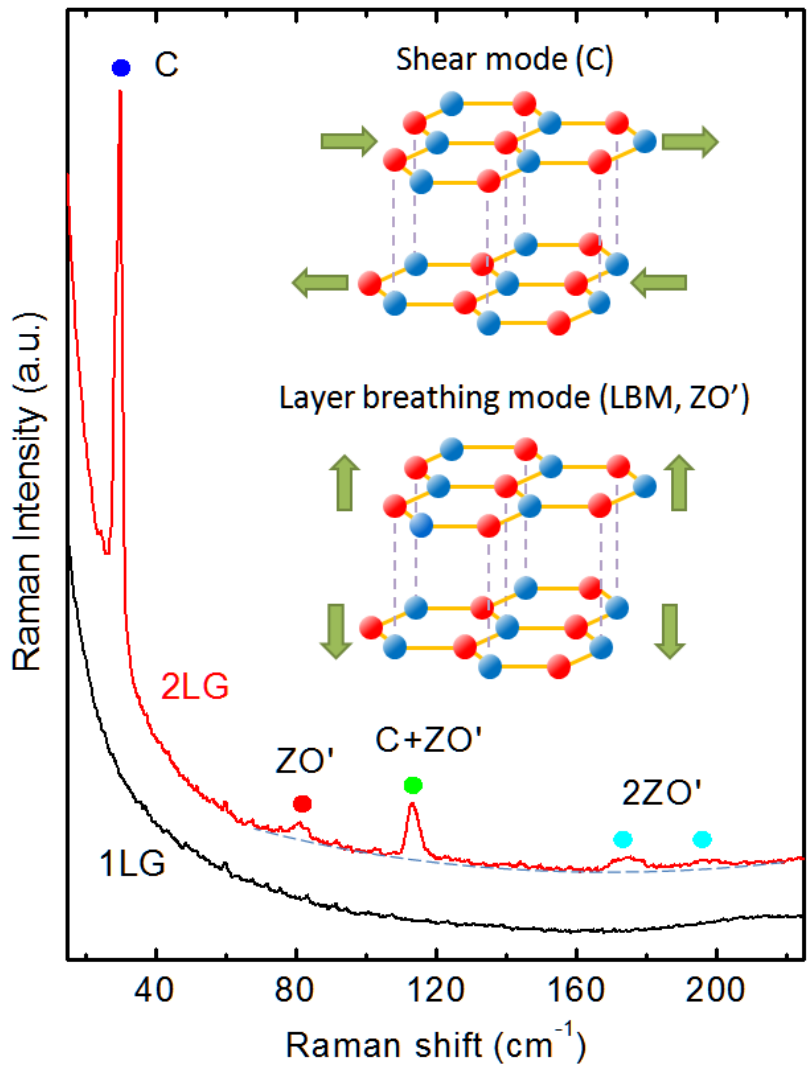

**Figure 1.** Low-frequency Raman spectra from monolayer graphene (1LG, black line) and bilayer graphene (2LG, red line). The blue and red dots denote, respectively, the shear (C) mode at 30 cm$^{-1}$ and the layer breathing mode (LBM or ZO') at 81 cm$^{-1}$. The green dot denotes the combination mode (C+ZO') at 113 cm$^{-1}$. The cyan dots denote the double peaks of the doubly resonant 2ZO' overtone mode. The dashed line highlights the background on which the weak Raman peaks superimpose. The insets are the schematic representations of the rigid-plane shear mode and LBM in 2LG.

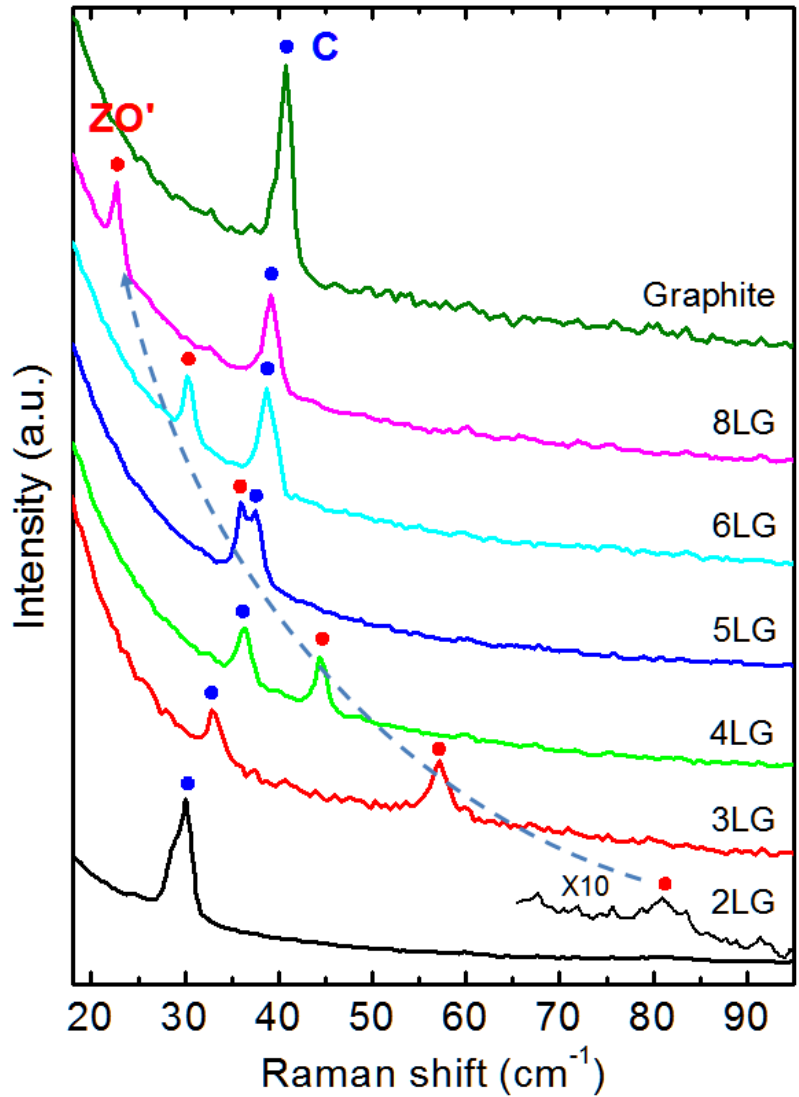

**Figure 2.** Low-frequency Raman spectra from *N*-layer graphene (*N*LG) with layer number *N* = 2 to 8 and from graphite. The blue dots denote the shear mode (C). The red dots denote the lowest-frequency layer-breathing (ZO') mode. The 2LG ZO' mode is multiplied by a factor of 10 for clarity. The arrow is a guide for the eye to highlight the redshift of the ZO' Raman mode.

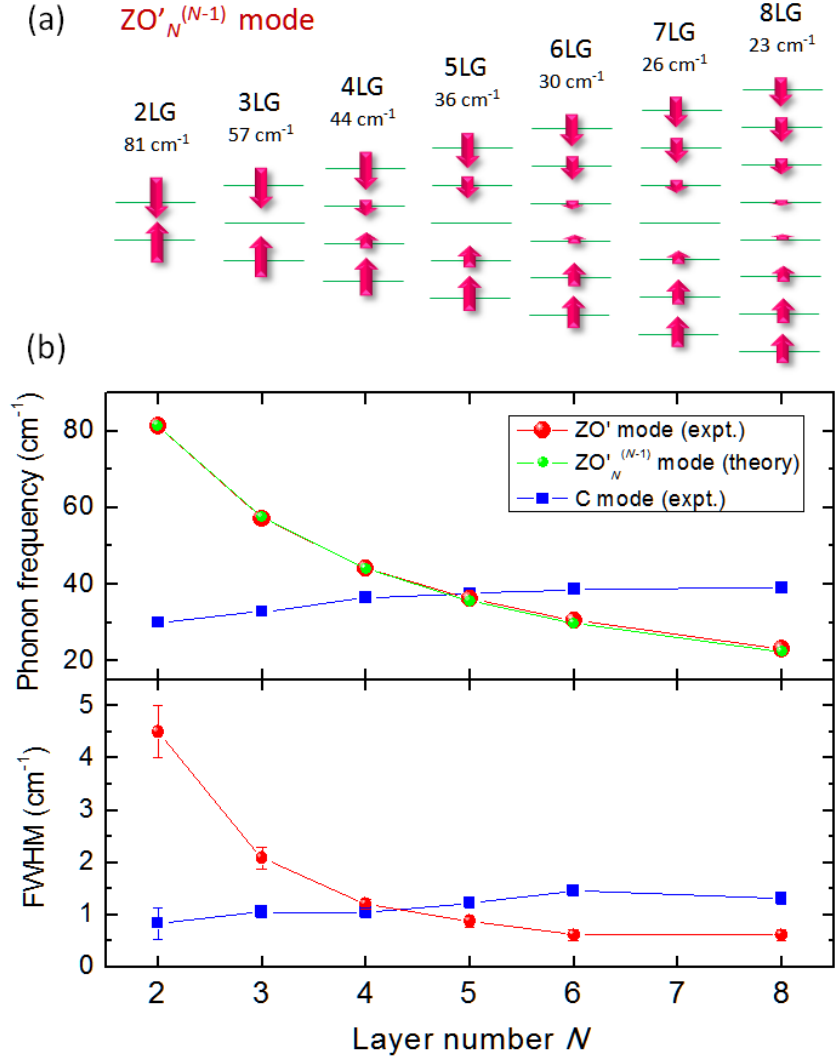

**Figure 3.** (a) Schematic layer displacement and measured vibrational frequencies for the lowest-frequency ZO'$_N^{(N-1)}$ modes in FLG (The frequency is calculated for 7LG). The green lines represent the graphene layers, and the red arrows represent the relative magnitude and direction of the layer displacement. (b) Peak frequency and full width at half maximum (FWHM) of the interlayer modes from Figure 2 as a function of layer number. The blue and red symbols are the experimental data for the C mode and ZO' mode, respectively. The green dots are the predicted frequency of the ZO'$_N^{(N-1)}$ mode based on a linear-chain model described in the text. The instrumental resolution (0.5 cm$^{-1}$) has been subtracted from the measured FWHM.

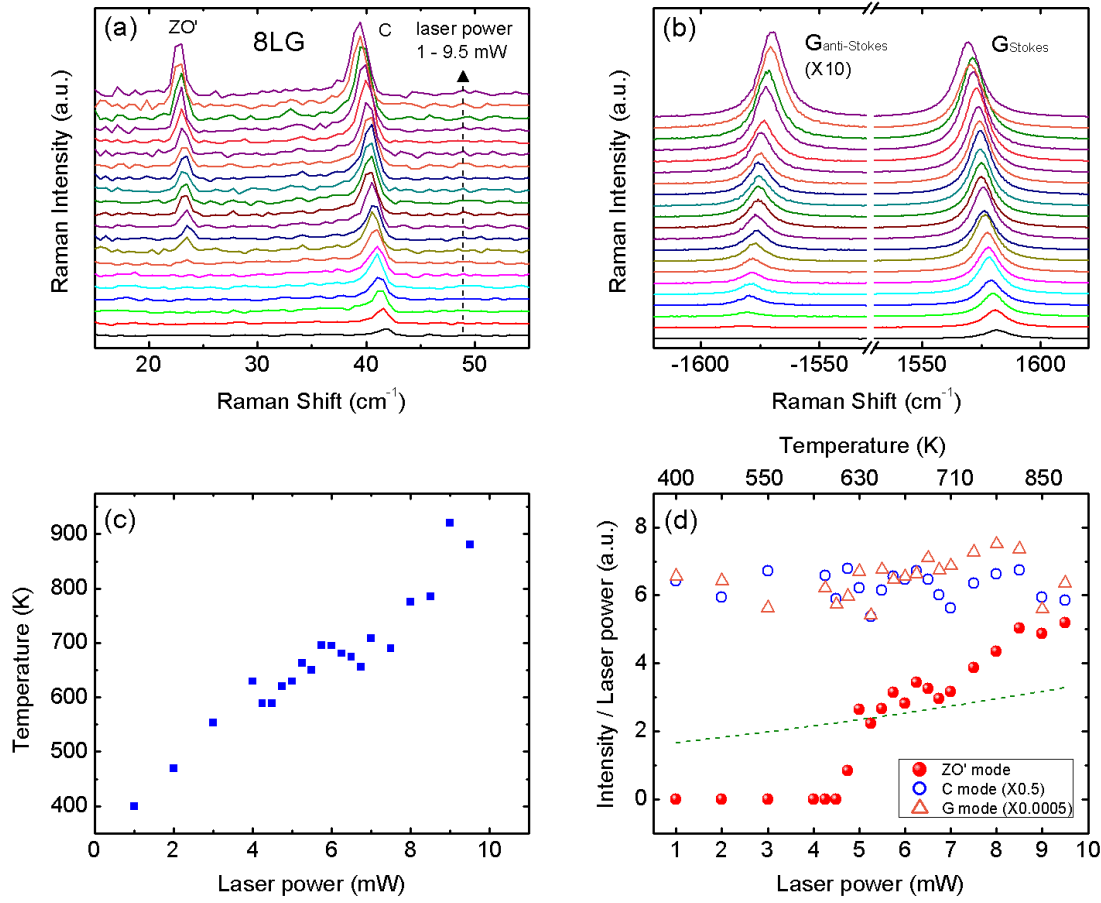

**Figure 4.** (a) Raman spectra of the layer breathing (ZO') and shear (C) modes from suspended 8LG under different excitation laser power. The spectra are background-subtracted and displaced vertically for clarity. From bottom to top, the spectra correspond to incident laser power of 1, 2, 3, 4, 4.25, 4.5, 4.75, 5, 5.25, 5.5, 5.75, 6, 6.25, 6.5, 6.75, 7, 7.5, 8, 8.5, 9 and 9.5 mW. (b) Raman spectra of anti-Stokes and Stokes G modes from 8LG at the respective laser power in Panel (a). The anti-Stokes peaks are multiplied by a factor of 10. (c) The graphene temperature extracted from the intensity ratio of Stokes to anti-Stokes G peaks in Panel (b). Instrumental efficiency has been taken into account. (d) Normalized integrated Raman intensities of the ZO', C and G modes as a function of the laser power (bottom axis) and temperature (top axis). The temperature is estimated from Panel (c) and is scaled slightly nonlinearly in the top axis. The C-mode and G-mode intensities are multiplied by a factor of 0.5 and 0.0005, respectively. The dashed line shows the expected variation of Raman intensity due to the increase of ZO' phonon population at increasing temperature.

# Supporting Information of
# Temperature-activated layer-breathing vibrations in few-layer graphene

Chun Hung Lui,[*,†,⊥] Zhipeng Ye,[‡,⊥] Courtney Keiser,[‡] Xun Xiao,[‡,∥] and Rui He[*,‡]

[†] Department of Physics, Massachusetts Institute of Technology, Cambridge, Massachusetts 02139, USA

[‡] Department of Physics, University of Northern Iowa, Cedar Falls, Iowa 50614, USA

[∥] College of Physical Science and Technology, Yangtze University, Jingzhou, Hubei 434023, China

[⊥] These authors contributed equally to this work.
*Corresponding author (email: lui@mit.edu; rui.he@uni.edu)

## 1. Supplemental tables

|  | 2LG | 3LG | 4LG | 5LG | 6LG | 7LG | 8LG |
|---|---|---|---|---|---|---|---|
| ZO'$_N^{(N-1)}$ | 81, $A_{1g}$, R | 57, $A_1'$, R | 44, $A_{1g}$, R | 35, $A_1'$, R | 30, $A_{1g}$, R | 25, $A_1'$, R | 22, $A_{1g}$, R |
| ZO'$_N^{(N-2)}$ |  | 99, $A_1''$, IR | 81, $A_{2u}$, IR | 67, $A_2''$, IR | 57, $A_{2u}$, IR | 50, $A_2''$, IR | 44, $A_{2u}$, IR |
| ZO'$_N^{(N-3)}$ |  |  | 106, $A_{1g}$, R | 93, $A_1'$, R | 81, $A_{1g}$, R | 71, $A_1'$, R | 64, $A_{1g}$, R |
| ZO'$_N^{(N-4)}$ |  |  |  | 109, $A_2''$, IR | 99, $A_{2u}$, IR | 90, $A_2''$, IR | 81, $A_{2u}$, IR |
| ZO'$_N^{(N-5)}$ |  |  |  |  | 111, $A_{1g}$, R | 103, $A_1'$, R | 95, $A_{1g}$, R |
| ZO'$_N^{(N-6)}$ |  |  |  |  |  | 112, $A_2''$, IR | 106, $A_{2u}$, IR |
| ZO'$_N^{(N-7)}$ |  |  |  |  |  |  | 112, $A_{1g}$, R |

**Table S1.** The predicted frequency (in units of cm$^{-1}$), symmetry group and optical activity of all the rigid-plane layer breathing modes (ZO'$_N^{(n)}$) in $N$-layer graphene ($N$LG) with Bernal stacking. The frequency is calculated based on the linear-chain model described in the main paper. The FLGs with odd number of layers have mirror crystal symmetry. A' (A'') denotes the irreducible representation of normal modes with even (odd) parity under mirror reflection. The FLGs with even layer number have inversion crystal symmetry. $A_{1g}$ ($A_{1u}$) denotes the representation of normal modes with even (odd) parity under inversion. The ZO'$_N^{(n)}$ modes are Raman-active (R) when $N - n$ is an odd number and are infrared-active (IR) when $N - n$ is an even number for all FLG systems. The symmetry analysis was adapted from Ref. [S1].

|  | 2LG | 3LG | 4LG | 5LG | 6LG | 7LG | 8LG |
|---|---|---|---|---|---|---|---|
| $C_N^{(N-1)}$ | 30, $E_g$, R | 21, E", R | 16, $E_g$, R | 13, E", R | 11, $E_g$, R | 9, E", R | 8, $E_g$, R |
| $C_N^{(N-2)}$ |  | 37, E', R+IR | 30, $E_u$, IR | 25, E', R+IR | 21, $E_u$, IR | 18, E', R+IR | 16, $E_u$, IR |
| $C_N^{(N-3)}$ |  |  | 39, $E_g$, R | 34, E", R | 30, $E_g$, R | 26, E", R | 24, $E_g$, R |
| $C_N^{(N-4)}$ |  |  |  | 40, E', R+IR | 37, $E_u$, IR | 33, E', R+IR | 30, $E_{2u}$, IR |
| $C_N^{(N-5)}$ |  |  |  |  | 41, $E_g$, R | 38, E", R | 35, $E_g$, R |
| $C_N^{(N-6)}$ |  |  |  |  |  | 41, E', R+IR | 39, $E_u$, IR |
| $C_N^{(N-7)}$ |  |  |  |  |  |  | 42, $E_g$, R |

**Table S2.** The predicted frequency (in units of cm$^{-1}$), symmetry group and optical activity of all the rigid-plane shear modes ($C_N^{(n)}$) in $N$-layer graphene (*N*LG) with Bernal stacking. As both the LBM and shear modes can be described by the linear-chain model, we calculated the shear-mode frequency similarly using Eq. (1) in the main paper based on the measured shear-mode frequency in 2LG (30 cm$^{-1}$). The FLGs with odd number of layers have mirror crystal symmetry. E' (E") denotes the irreducible representation of normal modes with even (odd) parity under mirror reflection. The FLGs with even layer number have inversion crystal symmetry. $E_g$ ($E_u$) denotes the representation of normal modes with even (odd) parity under inversion. "R" and "IR" denote Raman-active and infrared-active, respectively. The symmetry analysis was adapted from Ref. [S1].

## 2. Supplemental figures

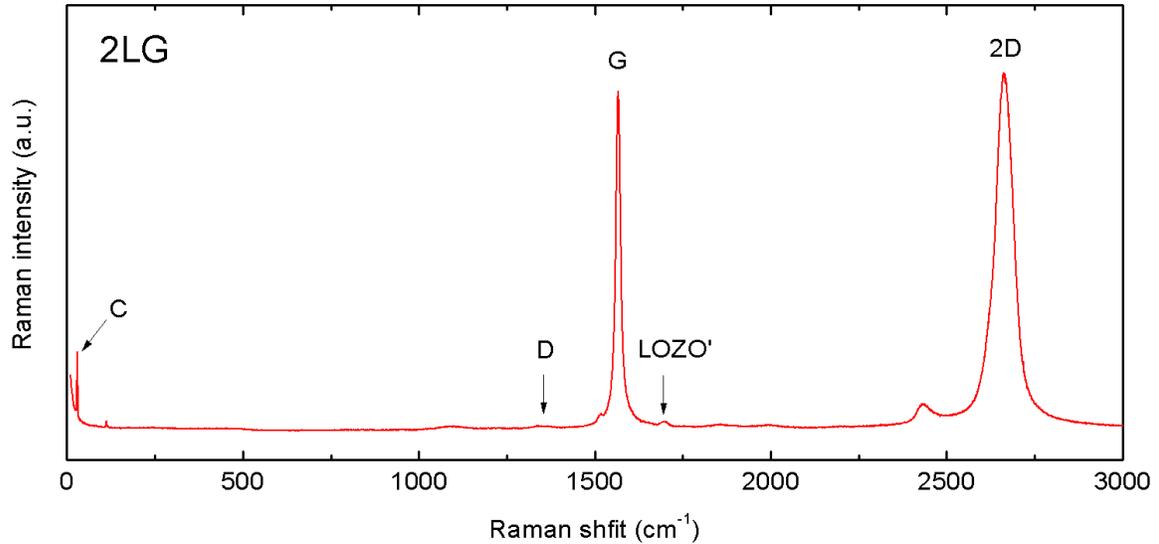

**Figure S1.** Raman spectrum of a suspended bilayer graphene (2LG) in a wide frequency range of 10 – 3000 cm$^{-1}$. The measurement was made with a 532-nm excitation laser at an incident power of ~10 mW in an argon-purged environment. Both the G and 2D modes exhibit significant redshift and thermal broadening due to the laser heating. No significant D band was observed, indicating the good sample quality under strong laser excitation.

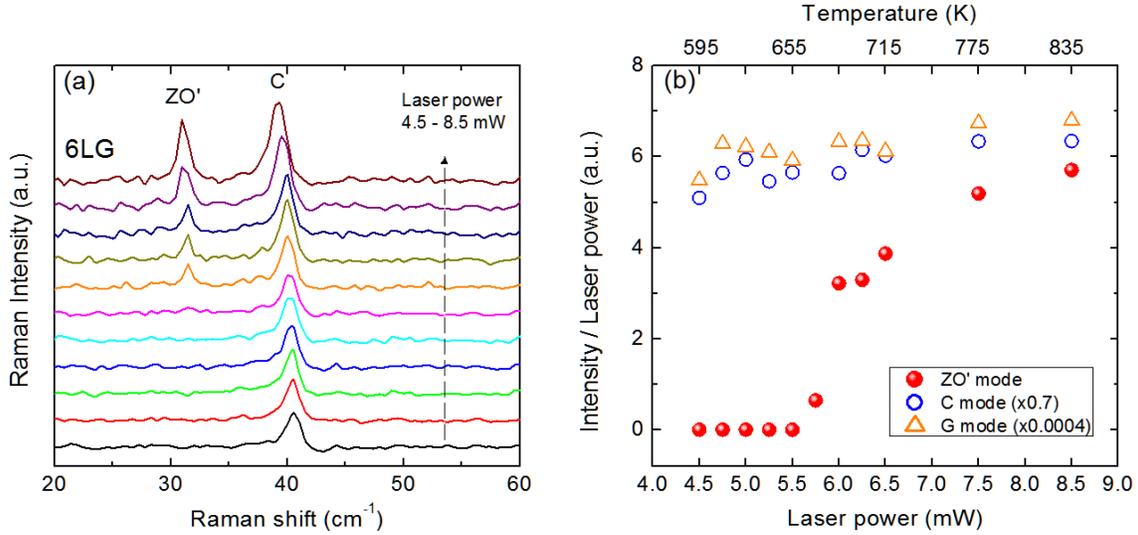

**Figure S2.** (a) Raman spectra of the layer breathing (ZO') and shear (C) modes from a suspended 6LG sample. The spectra are background-subtracted and shifted vertically for clarity. From bottom to top, the spectra correspond to incident laser power of 4.5, 4.75, 5, 5.25, 5.5, 5.75, 6, 6.25, 6.5, 7.5 and 8.5 mW. (b) Normalized integrated Raman intensities of the ZO', C and G modes as a function of laser power (bottom axis) and graphene temperature (top axis). The temperature is estimated from the intensity ratio of Stokes to anti-Stokes G peaks in each respective Raman measurement. The C-mode and G-mode intensities are multiplied by a factor of 0.7 and 0.0004, respectively. The onset temperature for the ZO' mode is approximately 670 K.

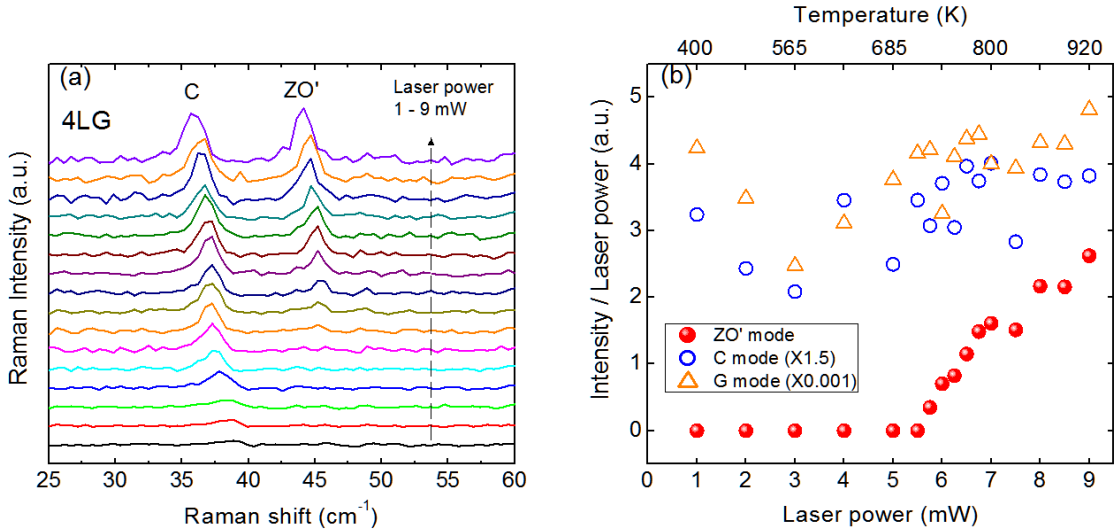

**Figure S3.** (a) Raman spectra of the layer breathing (ZO') and shear (C) modes from a suspended 4LG sample. The spectra are background-subtracted and shifted vertically for clarity. From bottom to top, the spectra correspond to incident laser power of 1, 2, 3, 4, 5, 5.5, 5.75, 6, 6.25, 6.5, 6.75, 7, 7.5, 8, 8.5 and 9 mW. (b) Normalized integrated Raman intensities of the ZO', C and G modes as a function of laser power (bottom axis) and graphene temperature (top axis). The graphene temperature is estimated from the intensity ratio of Stokes to anti-Stokes G peaks in each respective Raman measurement. The C-mode and G-mode intensities are multiplied by a factor of 1.5 and 0.001, respectively. The onset temperature for the ZO' mode is approximately 710 K.

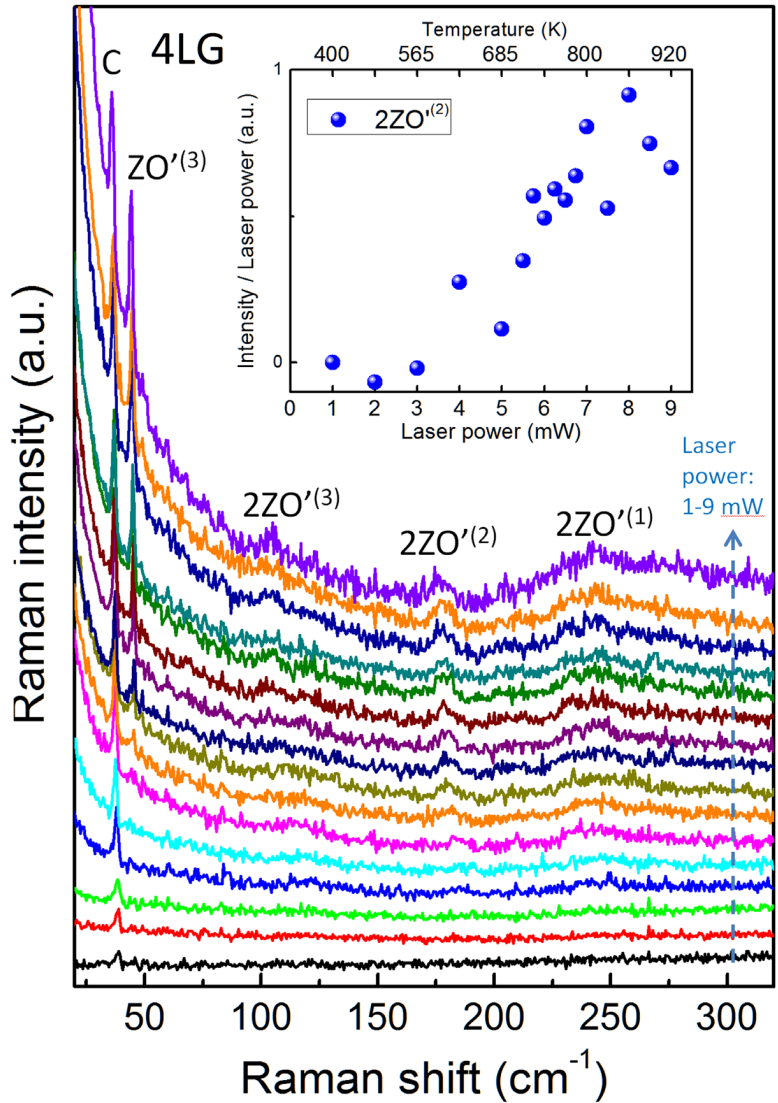

**Figure S4.** (a) The same Raman spectra as in Figure S3 for suspended 4LG in a wider spectral range without background subtraction. Besides the fundamental layer breathing mode of the lowest branch (ZO'$^{(3)}$) and the shear mode (C), the spectra also reveal the doubly resonant overtone LBMs of all three branches (2ZO'$^{(1)}$, 2ZO'$^{(2)}$ and 2ZO'$^{(3)}$ modes) in 4LG. The inset shows the integrated Raman intensity of the 2ZO'$^{(2)}$ mode as a function of excitation laser power (bottom axis) and the estimated graphene temperature (top axis). The onset temperature of the 2ZO'$^{(2)}$ mode (~630 K) appears to be lower than that of the fundamental LBM (~710 K) as shown in Figure S3b.

## Supplemental reference